\documentclass[lettersize,journal]{IEEEtran}
\usepackage{algorithmic}
\usepackage{array}
\usepackage[caption=false,font=normalsize,labelfont=sf,textfont=sf]{subfig}
\usepackage{cite}
\usepackage{amsmath,amssymb,amsfonts}
\usepackage{graphicx,color}
\usepackage{textcomp}
\usepackage{xcolor}
\usepackage{hyperref}
\hypersetup{hidelinks=true}
\usepackage{tabularx}

\usepackage{adjustbox}
\usepackage{makecell}
\usepackage{booktabs}

\usepackage{tikz}
\usepackage{tcolorbox}
\usepackage{courier}
\usepackage{minted}
\usepackage{amssymb}
\usepackage{pifont}
\newcommand{\cmark}{\color{blue}\checkmark} %
\newcommand{\xmark}{\color{red}\ding{55}}%
\def\BibTeX{{\rm B\kern-.05em{\sc i\kern-.025em b}\kern-.08em
    T\kern-.1667em\lower.7ex\hbox{E}\kern-.125emX}}
\usepackage{balance}

\begin{document}


\begin{titlepage}
    \centering
    \vspace*{\fill}

    {\Large
    This work has been submitted to the IEEE for possible publication.\\[1em]
    Copyright may be transferred without notice, after which this version may no longer be accessible.
    \par}

    \vspace*{\fill}
\end{titlepage}

\clearpage


\title{AUDDT: A Unified Benchmark Toolkit for Audio and Speech Deepfake Detectors}
\author{Yi Zhu, Heitor R. Guimarães, Arthur Pimentel, and Tiago Falk
\thanks{Yi Zhu was with INRS-EMT when this study was performed and is now with Reality Defender Inc.}%
\thanks{Heitor R. Guimarães was with INRS-EMT when this study was performed and is now with RBC Borealis.}%
\thanks{Arthur Pimentel, and Tiago Falk are with INRS-EMT and the INRS-UQO Joint Research Centre on Cybersecurity and Digital Trust.}%
}%

\markboth{Manuscript Under Review}%
{How to Use the IEEEtran \LaTeX \ Templates}

\maketitle


\begin{abstract}
With the prevalence of artificial intelligence (AI)-generated content, such as audio deepfakes, a large body of recent work has focused on developing deepfake detection techniques. However, existing benchmarks employ a narrow set of datasets, leaving detector generalization to real-world conditions uncertain. In this paper, we systematically review 31 existing audio deepfake datasets and present an open-source benchmarking toolkit called AUDDT (\url{https://github.com/MuSAELab/AUDDT}). The goal of this toolkit is to automate the evaluation of pretrained detectors across a wide range of speech and non-speech audio datasets, giving users direct feedback on the advantages and shortcomings of their deepfake detectors under diverse manipulation types and recording conditions. We start by showcasing the usage of the developed toolkit, the composition of our benchmark, and the breakdown of different deepfake subgroups. Next, we highlight how AUDDT differs from existing benchmarking efforts by enabling large-scale, diverse evaluation across modern spoofing methods and richer attribute-level analysis through comprehensive metadata annotation. Using a widely adopted pretrained deepfake detector, we present in- and out-of-domain detection results, revealing notable performance variability across different conditions and audio manipulation types. Lastly, we also analyze the limitations of these existing datasets and their gaps relative to practical deployment scenarios.
\end{abstract}


\section{INTRODUCTION}
\IEEEPARstart{T}{he} authenticity of digital audio is increasingly threatened by generative models capable of creating convincing synthetic speech, or `audio deepfakes'. The potential for this technology to be misused in sophisticated scams, impersonation attacks, and disinformation campaigns poses a significant risk to digital trust and security. To address this challenge, a large body of works has explored new detection systems to capture audio deepfakes~\cite{yi2023audio, zhang2025audio}.

While newly proposed models keep reporting state-of-the-art (SOTA) performance, their evaluation setups have typically relied on a handful of datasets curated in-lab, where the data diversity is far from that seen in real-world scenarios~\cite{chandra2025deepfake, jung2025spoofceleb}. Some representative widely-used datasets include the ASVspoof series (i.e., challenges hosted in 2019, 2021, 2024)~\cite{Nautsch_2021, yamagishi2021asvspoof2021, wang2026asvspoof}, the in-the-wild dataset~\cite{muller2022does}, and early versions of MLAAD~\cite{muller2024mlaad}, where performance saturation has started to be reported~\cite{yi2023audio, zhang2025audio}. Meanwhile, recent studies are reporting severe degradation when these SOTA pretrained detectors are evaluated on real-world deepfake media; in some cases, with accuracies dropping to near chance-levels~\cite{chandra2025deepfake}.

To bridge the gap between in-lab generated samples and real-world attacks, several new deepfake datasets have emerged recently, where data are generated using more advanced techniques (e.g., language models~\cite{wu2024codecfakeenhancingantispoofingmodels, chen2025codecfake+}, diffusion models~\cite{bhagtani2025diffssd, firc2024diffuse}) and with more conditions covered (e.g., new languages~\cite{muller2024mlaad} and accents~\cite{florez2023habla}, recording conditions~\cite{muller2025replay}, and perturbations~\cite{wang2026asvspoof}). Yet, new detection models continue to be primarily benchmarked on the conventional datasets, making it difficult to assess their performance against the full spectrum of modern deepfakes.

We argue that a primary reason for this disconnect is the lack of a unified and transparent benchmark for comprehensive model evaluation. The ASVspoof challenges~\cite{Nautsch_2021, yamagishi2021asvspoof2021, wang2026asvspoof}, for instance, were treated as a golden standard benchmark in early research because they provided a standardized evaluation protocol for fair model comparison. As more datasets emerge, however, the landscape of audio deepfake has become fragmented, as each dataset places focus on a certain aspect of deepfakes. For example, the most recent ASVspoof5 Challenge has introduced a wide variety of perturbations applied to both real and fake speech, such as codec compression, to evaluate model robustness~\cite{wang2026asvspoof}. 

Some other datasets such as CodecFake~\cite{wu2024codecfakeenhancingantispoofingmodels} and Codecfake~\cite{xie2024codecfakedatasetcountermeasuresuniversally} focus on model sensitivity to neural codec related artifacts where speech content remains intact. Another group of datasets, such as DiffuseOrConfuse~\cite{firc2024diffuse} and DiffSSD~\cite{bhagtani2025diffssd} are mainly used to gauge detectors' generalizability to diffusion based methods. Furthermore, datasets containing only vocoder-resynthesized speech or enhanced speech (e.g., \cite{frank2021wavefake}, \cite{li2024safeear}, \cite{guimaraes2025ditse}) blur the line between benign and malicious manipulations. As a consequence, such fragmentation makes it difficult to conduct fair model comparisons and establish robust baselines, as there is no standard protocol for evaluation across this diverse set of conditions.

More recently, some benchmarks have been proposed to partially address this issue. For example, Speech DF Arena~\cite{dowerah2026dfarena} introduces a large-scale benchmark and leaderboard to study evaluation generalization across out-of-domain spoofing conditions, while DeepFense~\cite{kheir2026deepfense} provides a reproducible framework for systematic analysis of model robustness under diverse training and evaluation configurations. Spoof-SUPERB~\cite{hashim2026spoofsuperb} further standardizes the evaluation of self-supervised speech representations for spoof detection across multiple downstream tasks. Despite these advances, current benchmarks differ substantially in their dataset coverage, annotation granularity, and support for unified cross-dataset evaluation protocols.

In this paper, we propose {\bf AUDDT} (pronounced as `audit'), a unified audio deepfake detection benchmark toolkit, which allows any pretrained model to be evaluated across a wide array of datasets with minimal manual effort to the end user. To simulate diverse, near real-world conditions, AUDDT incorporates 31 datasets that covers different real and fake speech categories (e.g., different generative methods, with/without perturbations, usage of vocoders or neural codec models, diverse languages/accents, etc.), which enables detailed metrics obtained for each category. Using AUDDT, we benchmark a baseline audio deepfake detection model pretrained on \textsc{ASVspoof2019}, and report the large variance in performance seen across the different data categories. This corroborates the importance of a unified, comprehensive benchmark to explore model weaknesses. Additionally, we discuss the limitations of these existing datasets and highlight their gaps with real-world conditions.

Compared to existing benchmarks, AUDDT defines a unified and extensible evaluation scheme for audio deepfake detection, providing more comprehensive coverage and interpretable analysis than existing benchmarking efforts. First, it increases both the diversity and scale of evaluation by providing a larger collection of datasets, including modern generative, codec-based, and in-the-wild manipulations, as well as additional non-speech audio categories, such as EnvSSD~\cite{yin2025envssd}, FakeSound~\cite{xie2024fakesound} and VCapAV~\cite{wang2025vcapav}. Second, AUDDT allows for grouped benchmarking through richer metadata annotations, enabling fine-grained analysis across multiple factors such as attack type, generation pipeline, environmental conditions, and domain characteristics, rather than relying solely on dataset-level aggregation. Finally, AUDDT is designed as a modular toolkit that can be easily integrated with existing evaluation or training pipelines, enabling seamless interoperability with models and protocols developed under other benchmarking frameworks. Together, these properties support more consistent, scalable, and interpretable evaluation of modern audio deepfake detection systems.

The remainder of this paper is organized as follows. Section~\ref{sec: existing works} describes existing audio deepfake datasets and the challenges seen without having a unified benchmarking toolkit. Section~\ref{sec: auddt} introduces the benchmark composition and a detailed usage of the proposed toolkit. Section~\ref{sec: results} presents the performance obtained with a pretrained baseline detector and discusses the limitations of existing datasets. Section~\ref{sec: conclusion} provides the conclusions.

\begin{figure*}
    \centering
    \includegraphics[width=\textwidth]{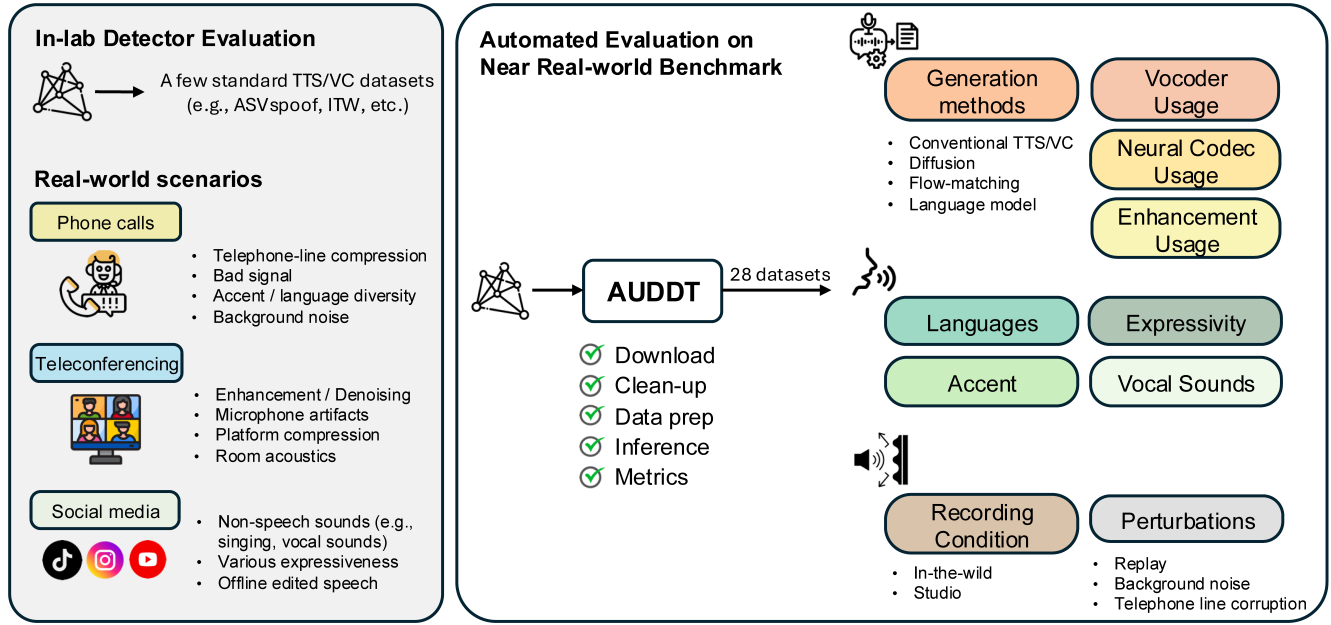}
    \caption{Functionality and coverage of AUDDT. Existing deepfake detectors are typically evaluated on a handful of datasets generated in a lab setting, such as the ASVspoof series, of which the performance does not truly reflect real-world performance. AUDDT offers automated benchmarking of any pretrained detector on a collection of 31 datasets, covering different types of real and deepfake speech that likely appear in real-world scenarios.}
    \label{fig:auddt}
\end{figure*}

\section{EXISTING DATASETS}
\label{sec: existing works}
\subsection{Landscape of Audio Deepfake Datasets}
Early deepfakes relied on conventional text-to-speech (TTS) and voice conversion (VC) models, where inputs are usually coded into hand-crafted features (e.g., phoneme sequence for text and mel-frequency cepstral coefficients, MFCC, for speech) then fused with disentangled speaker embeddings (e.g., d-vectors~\cite{wan2018generalized}), prosody embeddings, and estimated phoneme durations. The acoustic models were also smaller in size (e.g., using recurrent neural networks, RNNs) and usually relied on spectrograms or MFCCs as intermediate outputs. With these systems, a vocoder is therefore needed at the end to convert the intermediate outputs into the waveform. Some representative models that follow this type of design include variants of Tacotron~\cite{wang2017tacotron, skerry2018towards} and the FastSpeech series~\cite{ren2019fastspeech, ren2020fastspeech}, to name just a few. As a consequence, generated samples in earlier deepfake datasets (e.g.,~\cite{Nautsch_2021, reimao2019dataset, muller2022does}) typically have lower intelligibility and quality than real speech. 

Generative methods later improved on controllability via explicit modeling of other para-linguistic attributes, such as timbre, speed, and emotion~\cite{casanova2022yourtts, tan2024naturalspeech, jiang2023mega}, resulting in higher quality and improved naturalness deepfakes~\cite{muller2024mlaad, saito2024src4vc, salvi2023timit}. More recently, the adoption of more powerful acoustic model architectures, such as diffusion and flow-matching, have further improved generation quality~\cite{ju2024naturalspeech, du2024cosyvoice}. Due to their super-realistic outputs, datasets such as \textsc{DiffSSD}~\cite{bhagtani2025diffssd}, \textsc{DiffuseOrConfuse}~\cite{firc2024diffuse}, and \textsc{DFADD}~\cite{du2024dfadd} have been curated to study the generalizability of deepfake detectors pretrained with conventional TTS/VC generated samples. 

More recently, with the prevalence of large language models (LLMs), generative models have relied on fine-grained textual prompts as input, leading to additional gains in controllability and expressivity~\cite{du2024cosyvoice, wang2023neuralcodeclanguagemodels}. Since LLMs often require discrete tokens as input, neural codec models are typically used for audio encoding~\cite{mousavi2025discrete}. To study the characteristics of neural codec models, as well as whether the related artifacts can be used to recognize LLM-generated speech, datasets such as CodecFake~\cite{wu2024codecfakeenhancingantispoofingmodels} and Codecfake~\cite{xie2024codecfakedatasetcountermeasuresuniversally} have been developed.

While the aforementioned datasets focused on changes in deepfake generation methods, other datasets have explored the impact of perturbations likely encountered in real-world scenarios. For example, ASVspoof2021-LA~\cite{yamagishi2021asvspoof2021} and ASVspoof5~\cite{wang2026asvspoof} studied the impact of telephony corruptions (e.g., voice-over-internet-protocol, VoIP, and public switched telephone network, PSTN, systems) and codec compressions (e.g., mp3, VGG, Opus). Other datasets, such as FoR-rerecorded~\cite{reimao2019dataset} and replayDF~\cite{muller2025replay} have been developed to gauge model robustness to replay artifacts. In addition to these physical perturbations, a few datasets such as WaveFake~\cite{frank2021wavefake}, CVoiceFake~\cite{li2024safeear}, and LibriSeVoc~\cite{sun2023ai}, have tested the use of spectrograms of real speech input directly to different vocoders, leaving the speech content and paralinguistic attributes intact. While the resulting speech signal is not necessarily a `deepfake', these vocoder-generated samples can be used to study model sensitivity to vocoder-related neural artifacts.

\subsection{Gaps with Real-world Conditions}
Despite the wide coverage of existing deepfake datasets, as highlighted in the previous section, a large discrepancy is often seen when detectors are tested in real-world conditions~\cite{ge2025post, chandra2025deepfake}. This gap largely exists because real-world data contains layered sources of variation that are often absent in curated datasets (see left subplot in Figure.~\ref{fig:auddt}). Taking teleconferencing as an example, beyond the typical compression from codecs (e.g., Opus), the source audio is frequently processed by real-time denoising modules. A highly sensitive detector might misinterpret artifacts from such denoising as evidence of manipulation, causing it to misclassify real human speech. Additionally, variables such as microphone quality and room acoustics can alter speech characteristics, either by introducing their own artifacts or by masking subtle deepfake attributes. Similar issues exist for other scenarios, such as real-time phone calls or social media content, where various speech and acoustic conditions are intertwined, posing a challenge for model generalizability and robustness. 

To help overcome this gap, we argue that modern deepfake detectors need to be benchmarked on a large variety of deepfake datasets. A granular, category-by-category performance analysis is essential to identify specific vulnerabilities that may arise in real-world scenarios. To meet this need, we introduce AUDDT, a toolkit that enables automated and comprehensive evaluation for deepfake researchers.

\subsection{Existing Audio Deepfake Detection Benchmarks and Frameworks}

Recent efforts have sought to improve the standardization and reproducibility of audio deepfake detection benchmarking. DeepFense~\cite{kheir2026deepfense}, Speech DF Arena~\cite{dowerah2026dfarena}, and Spoof-SUPERB~\cite{hashim2026spoofsuperb} each address different aspects of this problem, including cross-dataset evaluation, fairness-aware benchmarking, and standardized evaluation of self-supervised speech representations. These frameworks have contributed toward more systematic evaluation protocols and broader dataset coverage compared to earlier single-dataset evaluations.

Despite these advances, existing frameworks still differ substantially in evaluation scope, metadata structure, and supported audio conditions. Table~\ref{tab:benchmarks} compares these benchmarks with AUDDT across several attributes. Compared to prior works, AUDDT incorporates a larger and more diverse collection of datasets, including both speech and non-speech audio manipulations. Furthermore, AUDDT emphasizes richer metadata annotations spanning 10 general attributes, enabling fine-grained grouped benchmarking across multiple axes such as perturbation conditions, language, accent, generative method, and neural codec usage. Finally, the modular structure of AUDDT facilitates integration with externally trained models and existing benchmarking pipelines, allowing unified evaluation across heterogeneous datasets and detector architectures.

\setlength{\tabcolsep}{1.5pt}
\begin{table}[]
\centering
\caption{Attribute comparison of different audio deepfake detection benchmarks}
\footnotesize
\begin{adjustbox}{max width=\textwidth}
\begin{tabularx}{\columnwidth}{cccX}
\hline
Benchmark & Datasets & \makecell{Non-Speech \\ Data} & \makecell{Metadata \\ Granularity} \\ \hline \hline
DeepFense~\cite{kheir2026deepfense} & 13 & \cmark & Fairness features: speaker gender, language, and perceptual audio quality \\ \hline
Speech DF Arena~\cite{dowerah2026dfarena}  & 14 & \xmark &  Attack types: neural audio codec/vocoder artifacts, end-to-end TTS, and in-the-wild  \\ \hline
Spoof-SUPERB~\cite{hashim2026spoofsuperb} & 8 & \xmark & Acoustic degradations: noise, reverb, codecs \\ \hline
\textbf{AUDDT} & 30+ & \cmark & 10 general attributes: Category, in-the-wild, perturbation, language, accent, vocal sounds, expressivity, vocoder, neural codec and generative method \\ \hline
\end{tabularx}
\end{adjustbox}
\label{tab:benchmarks}
\end{table}

\section{AUDDT: BENCHMARK COMPOSITION AND TOOLKIT USAGE}
\label{sec: auddt}
\subsection{Overview}
As shown in Fig.~\ref{fig:auddt}, AUDDT provides an interface that bridges any upstream pretrained deepfake detector with several downstream deepfake datasets. At the time of writing, AUDDT covers 31 datasets, as listed in Table~\ref{tab:dataset}, and further divides them into groups based on their different characteristics. This partitioning allows users to investigate model performance across categories, resulting in a more comprehensive model evaluation. A major goal of AUDDT is to alleviate manual efforts on the end-user, therefore an end-to-end benchmarking pipeline is built, covering the initial dataset downloading to the final metrics calculation and report preparation. In the following subsections, we detail the current benchmark composition, as well as presents examples of toolkit usage.

\subsection{Benchmark composition}

\setlength{\tabcolsep}{1.5pt}
\begin{table*}[]
\centering
\caption{Datasets included in the current version of AUDDT (ordered alphabetically). Most columns only have binary options, referring to the existence of certain attributes, such as the usage of vocoders or neural codec models, the inclusion of perturbations, etc. The `Generative method' column indicates the generic type of generative models used for crafting deepfakes. Abbreviations: SVS=singing voice synthesis, SVC=singing voice conversion, ALM=audio language model, SR=super-resolution, V2A=video to audio}
\begin{tabularx}{\textwidth}{ccccccccccccc}
\hline
Dataset & Category & In-the-wild & Perturbation & Lang & Accent & Vocal sounds & Expressivity & Vocoder & Neural codec & Generative method \\ \hline \hline
ASVspoof5~\cite{wang2026asvspoof} & Speech DF & \xmark & \cmark & EN & \xmark & \xmark  & \xmark & \cmark & \xmark & TTS+VC \\ \hline
ASVspoof2019\_LA~\cite{Nautsch_2021} & Speech DF & \xmark & \xmark  & EN & \xmark & \xmark  & \xmark & \cmark & \xmark & TTS+VC \\ \hline
ASVspoof2021\_DF~\cite{yamagishi2021asvspoof2021} & Speech DF & \xmark & \cmark & EN & \xmark & \xmark  & \xmark & \cmark & \xmark & TTS+VC \\ \hline
ASVspoof2021\_LA~\cite{yamagishi2021asvspoof2021} & Speech DF & \xmark & \cmark & EN & \xmark & \xmark  & \xmark & \cmark & \xmark & TTS+VC \\ \hline
CodecFake~\cite{wu2024codecfakeenhancingantispoofingmodels} & Recoded Real+Fake & \xmark & \xmark & EN & \xmark & \xmark  & \xmark & \xmark & \cmark & Codec+LLM \\ \hline
Codecfake~\cite{xie2024codecfakedatasetcountermeasuresuniversally} & Recoded Real+Fake & \xmark & \xmark & CH & \xmark & \xmark  & \xmark & \xmark & \cmark & Codec+LLM \\ \hline
CtrSVDD~\cite{zang2024ctrsvdd} & Fake Singing Voice & \xmark & \xmark & 2 & \xmark & \xmark  & \cmark & \cmark & \xmark & SVS+SVC \\ \hline
CVoiceFake~\cite{li2024safeear} & Vocoded Real & \xmark & \xmark & 5 & \xmark & \xmark  & \xmark & \cmark & \xmark & Vocoders only \\ \hline
DECRO~\cite{decro} & Speech DF & \xmark & \xmark & 2 & \xmark & \xmark  & \xmark & \cmark & \xmark & TTS+VC \\ \hline
DFADD~\cite{du2024dfadd} & Speech DF & \xmark & \xmark & EN & \xmark & \xmark  & \xmark & \cmark & \xmark & Diff+Flow \\ \hline
DiffSSD~\cite{firc2024diffuse} & Speech DF & \xmark & \xmark & EN & \xmark & \xmark  & \xmark & \cmark & \xmark & Diffusion \\ \hline
DiffuseOrConfuse~\cite{firc2024diffuse} & Speech DF & \xmark & \xmark & EN & \xmark & \xmark  & \xmark & \cmark & \xmark & Diffusion \\ \hline
EnhanceSpeech~\cite{guimaraes2025ditse} & Enhanced Real & \cmark & \xmark & $\geq$2 & \cmark & \xmark  & \cmark & \cmark & \cmark & Enhancement \\ \hline
EnvSSD~\cite{yin2025envssd} & Audio Event DF & \xmark & \xmark & N/A & N/A & N/A  & N/A & \cmark & \cmark & ALM+Diff \\ \hline
FakeSound~\cite{xie2024fakesound} & Audio Event DF & \cmark & \xmark & N/A & N/A & N/A  & N/A & \cmark & \xmark & Diff+SR \\ \hline
FoR-original~\cite{reimao2019dataset} & Speech DF & \xmark & \xmark  & EN & \xmark & \xmark  & \xmark & \cmark & \xmark & TTS+VC \\ \hline
FoR-2seconds~\cite{reimao2019dataset} & Speech DF & \xmark & \xmark  & EN & \xmark & \xmark  & \xmark & \cmark & \xmark & TTS+VC \\ \hline
FoR-norm~\cite{reimao2019dataset} & Speech DF & \xmark & \xmark  & EN & \xmark & \xmark  & \xmark & \cmark & \xmark & TTS+VC \\ \hline
FoR-rerecorded~\cite{reimao2019dataset} & Speech DF & \xmark & \cmark  & EN & \xmark & \xmark  & \xmark & \cmark & \xmark & TTS+VC \\ \hline
HABLA~\cite{florez2023habla} & Speech DF & \xmark & \xmark  & EN & \cmark & \xmark  & \xmark & \xmark & \xmark & TTS+VC \\ \hline
In-the-wild~\cite{muller2022does} & Speech DF & \cmark & \xmark & EN & \xmark & \xmark  & \xmark & \cmark & \xmark & TTS+VC \\ \hline
MLAAD-v5~\cite{muller2024mlaad} & Speech DF & \xmark & \xmark & $>$100 & \cmark & \xmark  & \xmark & \cmark & \xmark & TTS+VC \\ \hline
MSceneSpeech~\cite{yang2024mscenespeech} & Real & \xmark & \xmark & 2 & \xmark & \xmark  & \cmark & \xmark & \xmark & N/A \\ \hline
ODSS~\cite{odss} & Speech DF & \xmark & \xmark & 6 & \xmark & \xmark  & \xmark & \cmark & \xmark & TTS+VC \\ \hline
Playback\_attacks~\cite{shang2021audio} & Replayed Real & \xmark & \cmark & EN & \xmark & \xmark  & \xmark & \xmark  & \xmark  & N/A \\ \hline
SpoofCeleb~\cite{jung2025spoofceleb} & Speech DF & \cmark & \xmark & EN & \xmark & \xmark  & \xmark & \cmark & \xmark & TTS+VC \\ \hline
JVNV~\cite{xin2024jvnv} & Real & \xmark & \xmark & JA & \xmark & \cmark  & \cmark & \cmark & \xmark & N/A \\ \hline
SRC4VC~\cite{saito2024src4vc} & Speech DF & \cmark & \xmark & JA & \xmark & \xmark  & \xmark & \cmark & \xmark & TTS+VC \\ \hline
TIMIT-TTS~\cite{salvi2023timit} & Fake & \xmark & \xmark & JA & \xmark & \xmark  & \xmark & \cmark & \xmark & TTS+VC \\ \hline
VCapAV~\cite{wang2025vcapav} & Audio Event DF & \cmark & \xmark & N/A & N/A & N/A & N/A & \xmark & \xmark & Diff+V2A \\ \hline
WaveFake~\cite{frank2021wavefake} & Vocoded Real & \xmark & \xmark  & EN & \xmark & \xmark  & \xmark & \cmark & \xmark & Vocoders only \\ \hline
\end{tabularx}
\label{tab:dataset}
\end{table*}

Within AUDDT, the datasets are classified with 10 different attributes to highlight their unique features. These attributes include: \\
{\bf Category:} While the majority of the datasets contain both real (i.e., bonafide) and fake (i.e., spoof) speech samples, some datasets contain only real or only fake ones. For example, \textsc{Playback-attacks}~\cite{shang2021audio} dataset contains only the replayed version without a deepfake component. We therefore label them as `Replayed Real'. Meanwhile, \textsc{WaveFake}~\cite{frank2021wavefake} and \textsc{CVoiceFake}~\cite{li2024safeear} contain the real speech and their vocoder-generated version. While vocoder is a commonly used component in speech generation, using the vocoder alone cannot change speech content or paralinguistic attributes. Given the definition of deepfake, we label these two datasets as `Vocoded Real'. Apart from speech deepfake datasets, we further include \textsc{CtrSVDD}~\cite{zhang2025audio}, which is a singing voice deepfake dataset (`Fake Singing Voice'). Additionally, we incorporate a dataset of environmental, non-speech sounds, such as \textsc{EnvSSD}~\cite{yin2025envssd}, \textsc{FakeSound}~\cite{xie2024fakesound} and \textsc{VCapAV}~\cite{wang2025vcapav}. These are labeled as `Audio Event DF'.  

Lastly, to gauge the impact of speech enhancement on detection accuracy, given the prevalence of such methods in real-world speech applications, we include samples from three benchmark datasets used for evaluation in~\cite{guimaraes2025ditse}: DAPS, AQECC, and DEMO. For each dataset, we consider the original noisy recordings (and corresponding clean references, when available), along with their enhanced counterparts (denoted as `Enhanced real'). The enhanced samples were generated using five representative speech enhancement approaches: DiTSE~\cite{guimaraes2025ditse}, and SGMSE+~\cite{richter2023speech}, two state-of-the-art diffusion-based methods; Genhancer~\cite{genhancer}, a high-fidelity discrete token generation model; Miipher~\cite{koizumi2023miipher}, a restoration framework tailored for degraded speech; and HiFi-GAN-2~\cite{su2021hifi}, a conventional regression-based baseline.\\ 
{\bf In-the-wild:} Most deepfake datasets are generated from clean, studio-quality recordings (e.g., \textsc{LJspeech}~\cite{ljspeech17}). Models trained on studio-quality real and fake pairs inevitably suffer from degradation when tested on in-the-wild speech, which is usually delivered in a natural and conversational manner with overlapped environmental noise. We therefore highlight the ones that contain speech collected in-the-wild for users to gauge model robustness in non-studio scenarios, including \textsc{In-the-wild}~\cite{muller2022does}, \textsc{SpoofCeleb}~\cite{jung2025spoofceleb}, and \textsc{SRC4VC}~\cite{saito2024src4vc}. \\
{\bf Perturbation:} Commonly seen perturbations include (1) artifacts seen when voice is transmitted over the telephone line, such as packet loss, bitrate variations, and compressions; and (2) artifacts caused by audio replay, such as effects of different microphones and loudspeakers, as well as room acoustics. The former is represented by \textsc{ASVspoof2021-LA} and \textsc{ASVspoof5}, and the latter includes \textsc{ASVspoof2021-PA}, \textsc{FoR-rerecorded}, and \textsc{Playback-attacks}. It should be noted that some in-the-wild datasets may also carry perturbations (e.g., microphone artifacts of interviewed speech). However, since this information is not explicitly provided, they are excluded from the perturbation category. \\
\begin{figure*}
\centering
\includegraphics[width=\linewidth]{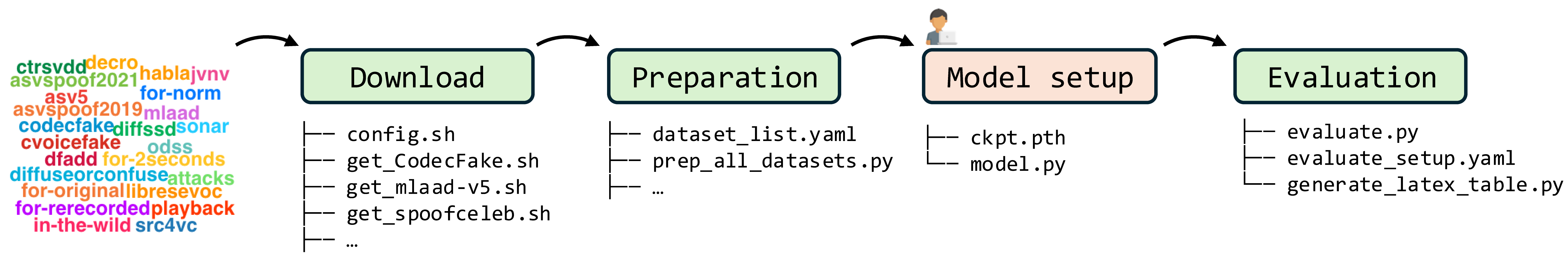}
\caption{Workflow of AUDDT. Data downloading, label standardization and preparation, as well as inference and metric calculation are automated. Users only need to provide a model architecture script as well as a checkpoint for benchmarking.}
\label{fig:workflow}
\end{figure*}
{\bf Language and accent:} Currently, deepfake datasets are dominated by the English language, with American and British accents. This is due to the fact that the majority of the open-source speech datasets have limited languages and accents. To gauge cross-lingual and cross-accent transferability of detectors, we include some Mandarin and Japanese datasets, along with a few multilingual datasets (e.g., \textsc{MLAAD-v5}~\cite{muller2024mlaad} and \textsc{CVoiceFake}~\cite{li2024safeear}). It should be highlighted that while \textsc{MLAAD-v5} has a wide coverage of languages, the speech for some languages are generated firstly via text translation then synthesized with TTS models, which differs from other datasets which do not involve a text translation step. \\
{\bf Expressivity and vocal sounds:} To evaluate performance on more expressive and emotional speech, we include datasets such as \textsc{MSceneSpeech}~\cite{yang2024mscenespeech} and \textsc{CtrSVDD}~\cite{zang2024ctrsvdd}. Because emotional expression often involves non-speech vocalizations, such as laughter, we also incorporate \textsc{JVNV}~\cite{xin2024jvnv}, which contains a wide variety of such sounds. \\
{\bf Usage of vocoder and neural codec models:} Many generative models rely on a final step to convert an intermediate representation (e.g., spectrogram) into a raw waveform. This is handled by either a vocoder or a neural codec model. To help isolate artifacts introduced by these components, our benchmark includes several specialized datasets. For traditional vocoders, datasets such as \textsc{WaveFake}~\cite{frank2021wavefake} and \textsc{CVoiceFake}~\cite{li2024safeear} provide pairs of real and `vocoded-real' speech. Similarly, for neural codecs --  which are essential for models that use discrete audio tokens -- datasets such as \textsc{CodecFake}~\cite{wu2024codecfakeenhancingantispoofingmodels} and \textsc{Codecfake}~\cite{xie2024codecfakedatasetcountermeasuresuniversally} offer pairs of real and `recoded-real' samples to study neural codec-specific artifacts. \\
{\bf Generative method:} Since each dataset incorporates several different generative model architectures, it is challenging to summarize all architectural components in one column. For simplicity, we highlight the main generative component (method) for each dataset. As mentioned in Section~\ref{sec: existing works}, the quality of generated speech has been improved significantly after diffusion, flow-matching, and LLMs are introduced. As such, we highlight those datasets crafted mainly using these generative methods. These datasets generally feature higher synthesis quality, posing a greater challenge to detection systems. Those incorporate more conventional TTS and VC methods are labeled as `TTS/VC'. For the vocoded, neural coded, and enhanced speech datasets, we mark them as `Vocoders only', `Codec', and `Enhancement'. Interested readers are encouraged to find dataset coverage details in the references.

\subsection{Toolkit Usage}
Next, we introduce the steps for benchmarking a pretrained detector using AUDDT. Figure~\ref{fig:workflow} depicts the workflow used and below we provide more details on each step.\\ 
{\bf Step-1: Data download:} When preparing our toolkit, we found data downloading and preparation to be a time-consuming process, as most datasets do not follow the same structure. For example, some datasets do not come with an accompanying label file and some do not include the source real speech used for deepfake generation. Therefore, as a first step we provide dataset-unique bash scripts that allow users to automatically download all datasets from the source links. Once users provide a target folder pointing to a centralized data storage location, all datasets will be pulled to the target folder without any manual efforts needed. \\
{\bf Step-2: Label preparation:} In the second step, all label files are standardized into the same format to be ingested by the downstream evaluation script. By default, each audio file is assigned a binary label of either `bonafide' or `spoof'. We recognize that some `bonafide' samples, such as those resynthesized by vocoders or neural codecs, may contain neural artifacts not present in original recordings. As such, some researchers may wish to treat these samples differently. To accommodate this, each dataset includes an independent preparation script, giving users the flexibility to customize the labels for their own needs. This entire step can be run with a single command.\\
{\bf Step-3: Pretrained model migration:} In this step, model scripts and pretrained checkpoints are needed from the users. The raw model script and checkpoint can be simply moved into the \texttt{models/} folder without any code changes. All raw models are then wrapped into a default \texttt{AudioDeepfakeDetector} class, so users can define customized functions to retrieve the probability score of a sample being fake. \\
{\bf Step-4: Evaluation:} Users can customize the evaluation setup via a single configuration \textsc{YAML} file (see Fig.~\ref{fig:yaml} for an example), based on which the subsequent inference and metrics calculation are automated. During benchmarking, users can either choose a single dataset to benchmark on or a certain group of datasets (e.g., `perturbed speech`). By default, all audios are resampled to 16kHz, as most audio and speech feature extractors are pretrained with 16kHz data. Meanwhile, since detectors are often trained with a fixed audio duration, the benchmarking can be performed with all audios aligned at the target duration (by either truncation or padding). In the end, the model probability scores will be saved per dataset along with the following metrics: equal error rate (EER), accuracy, true positive rate (TPR), true negative rate (TNR), and AUC-ROC. Optionally, users can choose to have results output in a \texttt{LaTeX} table and a PDF report with visualizations (e.g., bar plots).

It should be highlighted that we use a fixed threshold for calculating binary metrics, such as accuracy, TPR, and TNR. In the literature, some works choose the threshold obtained from the EER calculated on the test data to serve as the threshold for binarizing decisions. This means the threshold values will be different across test sets and the accuracy obtained can then be seen as an upper bound of the `true' accuracy without test information leakage. In real-world scenarios where evaluation data remain unseen, it is very challenging to have the threshold tailored accordingly, hence a fixed threshold needs to be set before deployment. While this threshold can be customized by users, a default value of 0.5 is used for reporting binary metrics.

\begin{figure}
\begin{minted}[
  frame=lines,      % Add a frame around the code
  framesep=2mm,     % Space between frame and code
  breaklines,
  % numbersep=2pt, 
  baselinestretch=1,
  fontsize=\small,  % Use a smaller font
  % linenos           % Add line numbers
]{yaml}
# Evaluation config file to interface upstream pretrained model with downstream datasets
model:
  path: models/detector_wrapper.py
  class_name: AudioDeepfakeDetector
  # Path to the checkpoint
  checkpoint: models/CHECKPOINT.pth
  device: 'cuda:0'
  model_args:
    # User's raw model script
    raw_model_path: models/baseline_model.py
    # Name of the top-level detector class
    raw_model_class_name: Model
    # Detector hyperparams can be set here
    raw_model_args:
      args: null
      model_device: 'cuda:0'

data:
  # Set this to a specific manifest.csv to benchmark on one dataset
  manifest_path: data/
  # Set to 'all' to benchmark on all datasets
  group_name: 'all'
  # Waveform preprocessing
  data_args:
    target_sample_rate: 16000 # resampling
    target_length: 64000 # duration limit (4s)

evaluation_settings:
  results_dir: results
  batch_size: 256 
  latex_output_path: results/examplar_table.tex
\end{minted}
\caption{An exemplar evaluation configuration file for benchmarking a baseline detection model on all datasets covered by AUDDT.}
\label{fig:yaml} 
\end{figure}

\subsection{Benchmark pretrained deepfake detector}
To highlight the usefulness of the toolkit, we employed a pretrained deepfake detector and used AUDDT to gauge its performance across the 31 different datasets. This detector (abbreviated as W2V-AASIST~\cite{tak2022automatic}) relies on Wav2vec2-XLSR-300M~\cite{babu2022xlsr} as the frontend and an AASIST classifier~\cite{tak21_asvspoof} as the backend, which has been finetuned solely on ASVspoof2019 training data. We chose this detector because it serves as a common baseline in the literature and its pretrained weights are publicly available, making it ideal for reproducibility. 

W2V-AASIST is a pretrained model that shares the architectural backbone of Wav2vec 2.0~\cite{baevski2020wav2vec20frameworkselfsupervised}, a framework composed of a convolutional feature encoder and a Transformer encoder. The convolutional encoder is composed of seven blocks of temporal convolution followed by layer normalization and a GELU activation layer.
The convolutional output representation is masked and passed to the Transformer as input. It employs a convolution-based relative position embedding layer with a kernel size of 128 and 16 groups at the bottom of the Transformer. The number of Transformer layers varies between different model versions. Here, our models utilize 24 Transformer encoder layers, totaling approximately 300 million parameters. Wav2vec2-XLS-R-300M is trained on 436 thousand hours of multilingual speech across 128 languages, enabling cross-lingual generalization~\cite{babu2022xlsr}

Usage of AUDDT requires migration of a pretrained detector, this means AUDDT-required libraries need to be installed within the detector environment. To avoid time spent on the users' side to address potential dependency conflicts, we intentionally kept minimal dependencies required for \mbox{AUDDT} and did not encounter any dependency conflicts when benchmarking W2V-AASIST. Following the original training preprocessing setup, all audios are resampled to 16kHz and amplitude normalized. The audio duration is aligned to 4s, where shorter audios are zero-padded while the first 4s are kept from longer audios. All benchmarking experiments are performed on Compute Canada with a single A100 instance.

\subsection{Evaluation metrics}

While most studies use equal error rate (EER) as the main evaluation metric, it can lead to an over-optimistic estimation of the real-world performance. This is because EER is defined as the arithmetic mean between the true positive rate (TPR) and false positive rate (FPR) evaluated at the EER threshold obtained per test dataset. For production deployment, however, a single threshold needs to be pre-determined for all test distributions. As such, EER values only reflect the upper bound of the discriminative power of detectors under different test conditions. We therefore choose accuracy as the main metric reported in the following findings, where a single threshold is used to binarize model scores. It is important to mention that our toolkit covers a wide range of metrics, including EER, accuracy, TPR, TNR, precision, AUC-ROC, F1 score, as well as the number of true/false positives and true/false negatives. For conciseness, we just report accuracy in the following sections.

\section{RESULTS AND DISCUSSIONS}
\label{sec: results}

\begin{figure*}
\centering
\includegraphics[width=\linewidth]{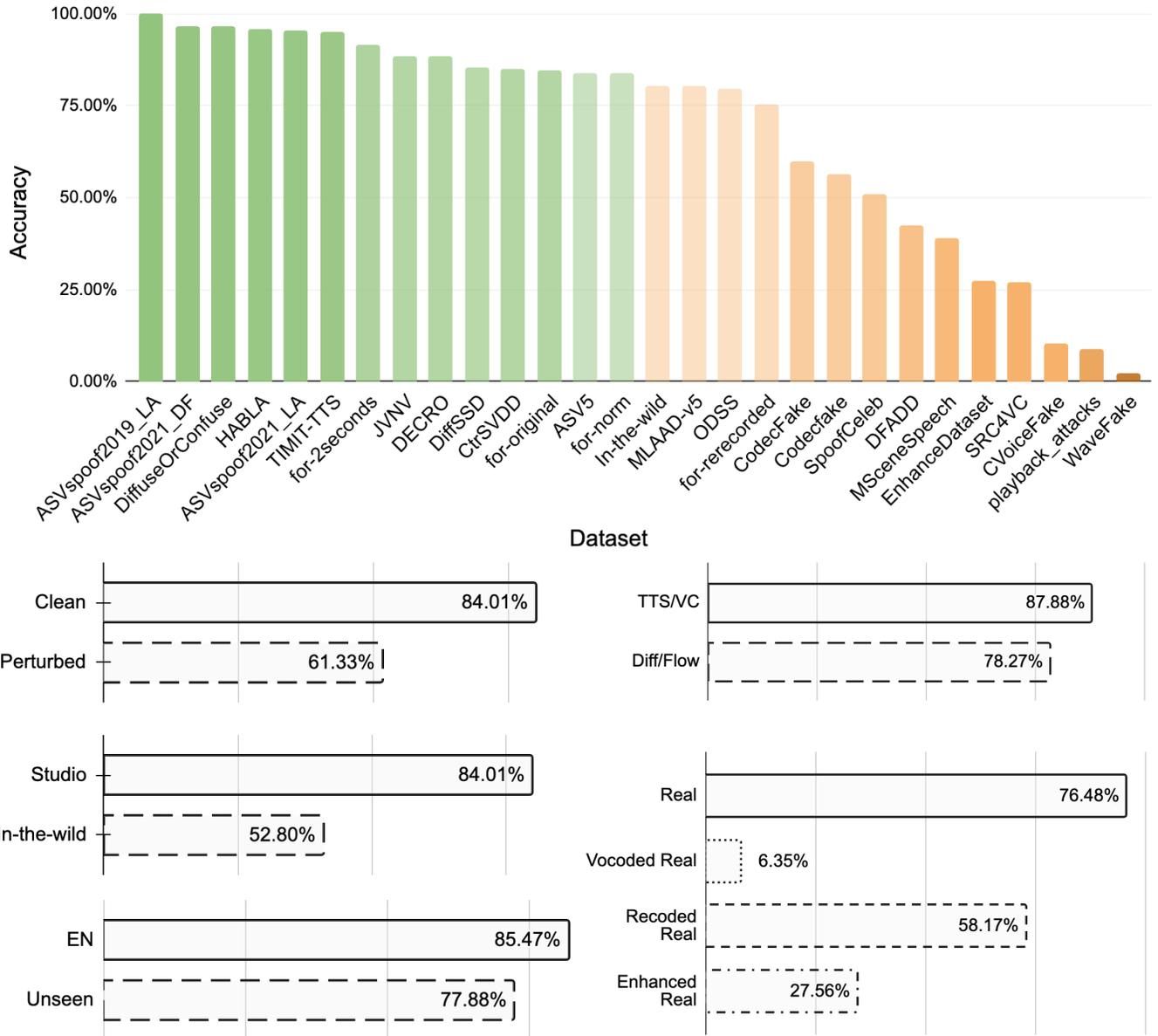}
\caption{Model performance (accuracy) on different datasets and groups. Across all datasets, the median accuracy is 83\%. Those with accuracies under the median are highlighted in orange. When computing group accuracies, vocoded, neural coded, and enhanced speech are excluded excepted for the bottom right subplot.}
\label{fig:performance}
\end{figure*}

\subsection{Benchmark results}
Figure~\ref{fig:performance} summarizes the performance obtained by the benchmark detector. Since some datasets only contain one class, we choose accuracy as the main metric. As can be seen, on standard deepfake datasets that have been widely employed, such as \textsc{ASVspoof2019-LA}, \textsc{ASVspoof2021-DF}, and \textsc{ASVspoof2021-LA}, W2V-AASIST achieves accuracy over 90\%. Similar performance is obtained on datasets that also rely on traditional TTS/VC methods, such as \textsc{TIMIT-TTS}. Furthermore, the baseline obtains 96\% accuracy on accented English (i.e., \textsc{HABLA}). These findings suggest that the benchmark detector can generalize to unseen attacks to some extent. 

On the other hand, a large performance variance is seen across datasets with higher-quality deepfakes, such as those crafted using diffusion model and flow-matching (i.e., \textsc{DiffSSD}, \textsc{DECRO}, \textsc{DiffuseOrConfuse}, and \textsc{DFADD}). While on \textsc{DiffuseOrConfuse} the accuracy can be as good as 96\%, on \textsc{DFADD} the accuracy drops to 42\%. Similar trends can be seen when the model is tested on unseen languages. For example, while the baseline model achieves 80\% and 79\% accuracy on the multilingual \textsc{MLAAD-v5} and \textsc{ODSS}, where English and other European languages are the major contributors, only 26\% accuracy is obtained with the Japanese dataset \textsc{SRC4VC}. This suggests that the learned deepfake patterns may transfer across closely related languages, but fail to extend to genetically unrelated ones. 

Meanwhile, on in-the-wild speech datasets, such as \textsc{SpoofCeleb}, model accuracy drops to near chance-level, suggesting a large distribution shift between studio-quality and in-the-wild speech. Regarding perturbations, the baseline model demonstrates decent robustness to some groups of corruptions, such as the telephony ones included in \textsc{ASVspoof2021-LA}. This is likely due to the augmentations performed during finetuning (e.g., RawBoost). However, more severe performance degradation is observed when the model is tested on a larger variety of perturbations (e.g., \textsc{ASVspoof5}), when bitrate, compressions, and neural codec artifacts are combined. Besides telephony perturbations, replay attacks are also found to trigger markedly higher error rate, where the accuracy drops to only 8.7\% on \textsc{Playback-attacks}. 

We further investigate the performance of the baseline model on vocoder and neural codec generated speech. As there is no change in linguistic and paralinguistic attributes, we expect detectors to be robust to these neural artifacts and to label resynthesized speech as real. On the contrary, the detector achieves accuracy as low as 2\% on \textsc{WaveFake}, 10\% on \textsc{CVoiceFake}, 56\% accuracy on \textsc{Codecfake}, and 60\% accuracy on \textsc{CodecFake}. These numbers suggest that existing detectors likely use vocoder or neural codec related artifacts as a shortcut for classifying deepfakes. While previous works have reported that training on vocoder-resynthesized speech can lead to better performance~\cite{li2024safeear}, we argue that this will lead to increased false positive rates when real speech is input to `unharmful' neural processors, such as the latest speech enhancement~\cite{douglas2024enh} and speech separation methods~\cite{li2025advancesspeechseparationtechniques}. The performance obtained on \textsc{Enhance-Speech} corroborates this argument. The detector misclassifies 82\% of enhanced real speech as fake, demonstrating that current pretrained detectors are sensitive to neural artifacts. In the most ideal case, a detector should accurately capture deepfakes while being robust to these `unharmful' neural artifacts. 

Additionally, we explore the performance achieved with singing voice deepfakes. With \textsc{CtrSVDD}, the baseline detector is able to achieve 85\% accuracy without being exposed to singing voice data during finetuning nor pretraining. This demonstrates the potential overlap between speech deepfake and singing deepfake cues.

Lastly, we investigate the cross-domain generalizability where the speech pretrained detector is tested on audio event deepfakes. As can be seen from Table.~\ref{tab:audio-events-results}, an average accuracy of 51.69\% is obtained across the three audio event deepfake datasets (i.e., \textsc{FakeSound}, \textsc{EnvSSD}, and \textsc{VCapAV}), which is markedly lower than the average accuracy obtained on speech deepfake datasets (70.73\%). Results here suggest that the deepfake traces learned from speech datasets are heavily conditioned on speech characteristics, which can only be partially transferrable to audio event deepfakes, even when audios are generated using similar generative model architectures. 

\begin{table*}
  \centering
  \caption{Baseline speech model performance on audio event deepfake datasets. EER and AUC are undefined for FakeSound (spoof-only dataset). The model was trained on ASVspoof LA (speech deepfakes) and evaluated zero-shot on audio events.}
  \label{tab:audio-events-results}
  \begin{tabular}{cccccccc}
    \toprule
    Dataset & EER (\%) $\downarrow$ & AUC $\uparrow$ & Acc (\%) $\uparrow$ & TPR (\%) $\uparrow$ & TNR (\%) $\uparrow$ & Pre (\%) $\uparrow$ & F1 $\uparrow$ \\
    \midrule
    FakeSound   & --    & --     & 39.87 & 39.87 & --    & --    & --     \\
    EnvSDD      & 42.70 & 0.5902 & 32.70 & 25.63 & 82.02 & 90.87 & 0.3999 \\
    VCapAV-dev1 & 57.79 & 0.4495 & 37.36 & 30.74 & 57.23 & 68.32 & 0.4240 \\
    VCapAV-dev2 & 51.46 & 0.5060 & 44.45 & 38.31 & 56.73 & 63.91 & 0.4791 \\
    VCapAV-dev3 & 54.80 & 0.4691 & 37.67 & 33.86 & 56.73 & 79.65 & 0.4752 \\
    \midrule
    Average     & 51.69 & 0.5037 & 38.05 & 32.14 & 63.18 & 75.69 & 0.4445 \\
    \bottomrule
  \end{tabular}
\end{table*}

\subsection{Limitations of existing datasets}
This benchmarking exercise has allowed us to also gauge the limitations of existing deepfake datasets. First, we observe a dominance of studio-quality, scripted speech. Our analysis of the 31 datasets shows that only 3 (around 10\%) are sourced `in-the-wild', and fewer than a quarter systematically include realistic perturbations, such as telephony compressions. The main reason for this is that fake samples are predominantly generated from clean speech, such as \textsc{LJspeech}~\cite{ljspeech17} or \textsc{VCTK}~\cite{Yamagishi2019CSTRVC}. Since these source datasets are originally curated for speech recognition and synthesis, the speech signals are collected in a controlled and indoor setting, which shares different characteristics than in-the-wild speech. As a consequence, detectors finetuned on clean deepfakes suffer performance deterioration when tested on noisy, unpredictable audios found on platforms, such as as social media or VoIP calls.

Another key limitation observed is the lack of diversity in terms of human expressions. For instance, among the speech deepfake datasets, only two of the speech datasets (7\%) explicitly cover different accents, which risks creating detectors that are biased and less effective for a global user base. Similarly, while real speech can be expressive, only three datasets (10\%) focus on emotional content, and just a single dataset is dedicated to non-speech vocal sounds, including laughter. This can cause detectors to mistake natural human expressivity as deepfake artifacts, leading to lower TNR.

Finally, this benchmarking exercise has highlighted that datasets are not keeping up with the rapid evolution of generative models. While some newer datasets include methods such as diffusion, the development of evaluation data lags behind. This observation also aligns with the findings presented in \cite{chandra2025deepfake}, where in-lab trained detectors failed on deepfakes generated from commercial platforms, likely due to differences in generative model architectures used. 

\subsection{Limitations of the benchmark and future work}
We recognize that deepfakes are a rapidly changing field and that newer datasets have emerged since this benchmarking exercise was initiated (e.g., \cite{chandra2025deepfake}, \cite{muller2025replay}) and which are not included here. The rapid pace of research in this domain means that any benchmark study will inevitably be just a snapshot in time. Our ongoing work focuses on expanding AUDDT to address these gaps, by incorporating newly-released datasets, as well as making AUDDT into a dynamic ``living benchmark'' sustained by community contributions of new datasets through a standardized validation process. 

\section{Conclusion}
\label{sec: conclusion}
In this paper, we introduced AUDDT, an open-source toolkit for automated benchmarking across 31 diverse datasets. Our experiments with a pretrained baseline detector revealed significant performance disparities across different conditions, confirming that narrow testing can hide critical model weaknesses. By providing a unified framework for comprehensive testing, AUDDT aims to enable more transparent benchmarking and help bridge the gap between in-lab datasets and real-world scenarios. Our goal is to maintain AUDDT as a continuously relevant tool that helps the research community keep pace with the evolving landscape of audio deepfakes.




\bibliographystyle{IEEEtran}

\bibliography{ref.bib}

\end{document}